\documentstyle[times,11pt,twoside]{acta}

\input{epsf}

\SetPages{0}{0}

\SetVol{46}{1996}

\begin{document}
\newfont{\bb}{timesbi at 12pt}
\begin{Titlepage}

\Title{CPPA -- a New Hydrodynamical Code for Cosmological
Large-Scale Structure Simulations}

\Author{Andrzej~~K~u~d~l~i~c~k~i$^1$~~Tomasz~~P~l~e~w~a$^2$~~Micha{\l}~~ 
R~\'o~\.z~y~c~z~k~a$^{3,1}$}{$^1$Nicolaus Copernicus Astronomical Center, 
Bartycka~18, 00-716~Warszawa, Poland\\
e-mail:kudlicki@camk.edu.pl\\
$^2$Max-Planck-Institut f\"ur Astrophysik, Karl-Schwarzschild-Str.~1,
85740~Garching, Germany\\
e-mail:tomek@MPA-Garching.MPG.DE\\
$^3$Warsaw University Observatory, Al.~Ujazdowskie~4, 00-478~Warszawa,
Poland\\
e-mail:mnr@camk.edu.pl
}
\Received{August 14, 1996}
\end{Titlepage}
\Abstract{
We present a new Eulerian code able to follow the evolution of
large-scale structures in the Universe in the weakly nonlinear
regime. We compare test results with a N-body code and analytical
results.}{cosmology: large-scale structure of Universe -- methods: numerical}

\Section{Introduction} 
The Universe is homogeneous on scales larger than at least 100~Mpc. Therefore, 
any simulations intended to completely predict the development of structures 
ought to work with a dynamical range 10~{\rm kpc}--100~Mpc. A 
disadvantage of this fact is that enormous catalogues of galaxies are needed 
for a comparison of models of large-scale structure versus observations. The 
first sufficient catalogue will be the SDSS (Gunn and Knapp 1993), to come 
within a few years. On the other hand, a great advantage of examining very 
large scale structures is that density contrasts on these scales are 
relatively low (\ie weakly nonlinear, with ${\delta=\frac{\rho-
\overline{\rho}}{\overline{\rho}}<1}$), and various fields observed 
(convolved) with large windows can be worked on either with the use of 
analytical methods (like perturbation theory) or numerical simulations with 
simplified physics (\eg effects of pressure, radiation {\it etc} can be 
neglected). Even in the weakly nonlinear regime there are many features which 
have not been examined with analytical methods, \eg moments of projected 
velocities or redshift distortions. All of the above give a good reason for 
writing a new, fast code, able to follow the evolution of structure in various 
cosmological models. 

We present a new Eulerian code, which can be applied to linear and weakly 
non-linear evolutionary stages of dense regions of the Universe and to trace 
the voids ($N$-body codes are not good for describing low density regions). The 
code is tested and the results are compared with theoretical predictions, as 
well as with results obtained from modified AP$^3$M $N$-body code by Couchman 
(1991),(1994). 

\Section{Numerical Method}
\Subsection{Description}
Cosmological Pressureless Parabolic Advection (CPPA) is an Eulerian code, 
solving the dynamical equations of a self-gravitating pressureless fluid 
(Poisson, continuity and Euler). 

$$\frac{\partial^{2}\Phi}{\partial r^{i} \partial r^{i}} = 4\pi G \varrho
\eqno(1)$$
$$\frac{\partial \varrho}{\partial \tau} +
\frac{\partial \varrho u^{k}}{\partial r^{k}}
= 0
\eqno(2)$$
$$\frac{\partial \varrho u^{i}}{\partial \tau} +
\frac{\partial \varrho u^{i} u^{k} }{\partial r^{k}} +
\varrho \frac{\partial \Phi}{\partial r^{i}}
= 0
\eqno(3)$$

The code works with a fixed and uniform 3-D Cartesian grid in
cosmological frame.  Periodic boundary conditions are employed, which
do not introduce any systematic deviations at grid boundaries, and are
easy to implement for solving Poisson equation using Fourier methods.
The Poisson solver is adapted from a standard Helmholtz equation
solving routine available in {\em FISHPAK} library (Adams,
Schwarztrauber, and Sweet 1980).

The advection algorithm is based on the PPM method of Colella and
Woodward (1984) with pressure terms set to zero. We replaced the
nonlinear Riemann solver by a simple procedure which defines the
interface values as the upwind value equal to one of the effective
states averaged over domains of dependence in neighboring zones. Since
there is no sufficient justification for steepening of the density
jumps we did not use PPM steepening algorithm. Also, the whole set of
dissipation procedures is not needed for present application. The
temporal evolution is accurate to second order while the spatial
advection is formally third order accurate. The non-cosmological
version of the code has been widely tested and proved isotropic and
precisely solving Euler and Poisson equations.

To adapt the code for cosmological purposes we changed the variables
and coordinates in Eq.~(1)--(3) following the recipe
given by Gnedin (1995). Let $\tau$, $r^{i}$, $u^{i}$, $\varrho$,
$\Phi$ denote time, positions, velocities, density and gravitational
potential in physical units. We define new set of variables $t$,
$x^{i}$, $v^{i}$, $\rho$ and $\phi$ given by Eqs~(4)--(8):

\setcounter{equation}{3}
\begin{eqnarray}
{\rm d}\tau & = & \frac{a^2}{{\rm H}_0}{\rm d}t,\\
r^{i} & = & a L x^i,\\
u^{i} & = & \frac{da}{dt}Lx^{i} + \frac{L{\rm H}_0}{a}v^{i},\\
\varrho & = & \frac{3{\rm H}_0^{2}\Omega_0}{8\pi G a^3}\rho,\\
\Phi & = & \left(\frac{L {\rm H}_0}{a}\right)^{2}\phi -
      \frac{aL^{2}x^{i}x^{i}}{2}\frac{d^{2}a}{d\tau^2},
\end{eqnarray}
where $a$ is the scale factor, ${\rm H}_0$ is  the Hubble constant today,
$\Omega_0$  is the dimensionless density parameter  today, $G$ denotes
the gravitational constant  and $L$ is a dimensional parameter
denoting the computational box size. The density $\rho$ has been
scaled to ${\langle\rho\rangle=1}$, so ${\rho=1+\delta}$. In these coordinates,
assuming that scale factor changes with time according to
Friedmann-Robertson-Walker solution, equations Eq.~(1)--(3) are
$$\frac{\partial^2\phi}{\partial x^{i}\partial x^i}= 
\frac{3a\Omega_0}{2}\delta,
\eqno(9)$$
$$\frac{\partial \rho}{\partial t}+
\frac{\partial(\rho v^k)}{\partial x^k}=0,
\eqno(10)$$
and
$$\frac{\partial(\rho v^i)}{\partial t}+
\frac{\partial(\rho v^iv^k)}{\partial x^k}+
\rho \frac{\partial \phi}{\partial x^i}=0.\eqno(11)$$
The form of Eqs~(9)--(11) is very similar to
Eqs~(1)--(3), expressed in physical
coordinates, which makes it relatively easy to adapt any
hydrodynamical code to work in the cosmological frame. For a given
cosmological model, only the functions $a(t)$ and $\tau(t)$ have to
be evaluated numerically before the simulation is started. The
present version works with ${\Lambda=0}$, but it can be easily
modified to deal with models with a non-zero cosmological constant,
which requires only adding one more term to Eq.~(9).

\Subsection{Initial Conditions}
For cosmological simulations we assume a Gaussian distribution of the initial 
density contrast, $\delta$, with a given power spectrum and amplitude in the 
linear regime (${\langle \delta^2\rangle^\frac{1}{2} \leq 0.05}$). First, the 
initial density field $\delta(\vec{x})$ is set up in the form of its Fourier 
transform $\tilde{\delta}(\vec{k})$, with amplitudes according to a given 
power spectrum and phases chosen randomly. To keep the density field real we 
impose 
$$\tilde{\delta}(-\vec{k})=\tilde{\delta}(\vec{k})^{\star}.\eqno(12)$$
To prevent exceeding the Nyquist frequency, the spectra are multiplied by
$e^{-(k/k_{{\rm cut}})^{16}}\!\!,$ where $ k_{{\rm cut}} $ is smaller than $k_{Nq}$
(typically, we used ${k_{{\rm cut}} /k_{Nq}=0.5\ldots 0.}7$). Next, the
field is Fourier-transformed back to the real space. Simulations of
cosmological large-scale structure are started at early stages, when
the linear approximation in the perturbation theory (see \eg Peebles 1980,
\S 11) can be applied. According to this approximation
$$\delta=A_1 D_1(\tau)+A_2 D_2(\tau),\eqno(13)$$
where $D_1$ and $D_2$ denote the growing and decaying modes
respectively. We assume that our starting evolutionary stage is
advanced enough for not taking the decaying mode into account, \ie
$D_2(\tau_{\rm start}) \ll D_1(\tau_{\rm start})$. From the continuity
Eq.~(2) we find that the initial velocity $\vec{u}$ and density
contrast $\delta$ are then related as follows:
$$\frac{{\rm d}\delta}{{\rm d}\tau}+\frac{1}{a}\nabla\cdot\vec{u}=0.\eqno(14)$$
Hence,
$$\nabla\cdot\vec{u}=-a{\rm H}f\delta,\eqno(15)$$
where ${\rm H}$ is the current value of the Hubble's expansion parameter 
and $f$ is given by
$$f=\frac{a}{D_1}\frac{{\rm d}D_1}{{\rm d}a}=\frac{a}{D_1}
\frac{{\rm d}D_1}{{\rm d}\tau}\frac{1}{{\rm d}a/{\rm d}\tau}
=\frac{1}{{\rm H}}\frac{\dot{D_1}}{D_1}$$
and can be well approximated by
$$f\approx\Omega^{0.6}+\frac{\Omega_\Lambda}{70}\left(
1+\frac{1}{2}\Omega\right).$$
$\Omega$ denotes here the density parameter at $\tau_{\rm start}$,
and $\Omega_\Lambda = \Lambda/3{\rm H}^2$ is related 
to the cosmological constant.
The vector field $\vec{u}$ is then the gradient of a scalar field
$\psi$, defined by the Poisson equation
$$\triangle\psi=a{\rm H}f\delta.\eqno(16)$$
Therefore, given the density, the Poisson solver can be used for
evaluating the initial velocity field.

\Section{The AP$^3$M Code}
Our comparison code is the adaptive P$^3$M developed by Couchman
(1991),(1994). In this paper we will note only our changes to the code. Apart
of altering the program's I/O we made two major changes. First of
them was to introduce a CFL type evaluation of next time-step length.
The second one was to change the time coordinate as explained
below, and adapt the code to work in cosmological frame with ${\Omega
\neq1}$, which was not supported by the original version.
In comoving coordinates,
the equation of motion of $i$-th particle in the expanding Universe
reads
$$\frac{{\rm d}^2\vec{x}_i}{{\rm d}\tau^2}+
2\frac{\dot{a}}{a}\frac{{\rm d}\vec{x}_i}{{\rm d}\tau}
=-\frac{G}{a^3}\sum_{j\neq i}m_j\frac{\vec{x}_i-\vec{x}_j}
{|\vec{x}_i-\vec{x}_j|^3},\eqno(17)$$
where $\tau$ denotes the physical time and $a$ the scale factor. As
in CPPA, we will use another time variable, $t$, defined by
Eq.~(4). Eq.~(17) will transform to:
$$\frac{{\rm d}^2\vec{x}_i}{{\rm d}t^2}=
-a \frac{3\Omega_0L^3}{8\pi N}
\sum_{j\neq i} 
\frac{\vec{x}_i-\vec{x}_j}{|\vec{x}_i-\vec{x}_j|^3},\eqno(18)$$
where $L$ is the computational box size and all of the $N$ particles have mass 
equal $\frac{3{\rm H}_0^2\Omega_0L^3}{8\pi GN}$, which is taken to be the mass 
unit. It is worth mentioning that by introducing $t$ instead of physical time 
$\tau$ we have got rid of the cosmological drag term (in the original 
Couchman's version of the code time-stepping was performed at constant-length 
steps in $a^\alpha$, where $\alpha $ is an arbitrary real exponent, and the 
calculations were time-staggered due to presence of the drag term). 

\Section{Code Tests}
The CPPA code has been tested versus the modified AP$^3$M and analytical 
results (the latter including both exact solutions and approximate 
calculations of the perturbation theory). 

\Subsection{Zel'dovich Pancake Test}
Zel'dovich Pancake is the final stage of the evolution of a single-wave 
perturbation, resulting in formation of a caustic. It may be also regarded as 
an idealization of a wall between two voids. The chief advantage of the 
pancake problem is that for $\Omega_0 = 1$ 
there exists an exact analytical solution: 
$$\varrho(x,a)=R\left(1-\frac{a}{a_c}\cos(kx)\right)^{-1},$$ where $a_c$ is a 
parameter (the value of $a$ at which an infinitely dense caustic is formed), 
and ${R=\frac{\sqrt{a_c^2-a^2}}{a_c}}$ is the normalization factor. We 
performed this test on a $64^3$ grid, setting up 1, 2, 4, 8 and 16 waves with 
initial dispersion 0.003 at an epoch corresponding to the scale factor equal 
to $0.001$ of its present value. 1-D sections of the initial conditions are 
presented in Fig.~1. Fig.~2 shows the resulting pancakes at ${a=0.1}$. 
\begin{figure}[htb]
    \protect\centerline{
    \hfill
    \epsfysize=14cm\epsffile[5 150 585 700]{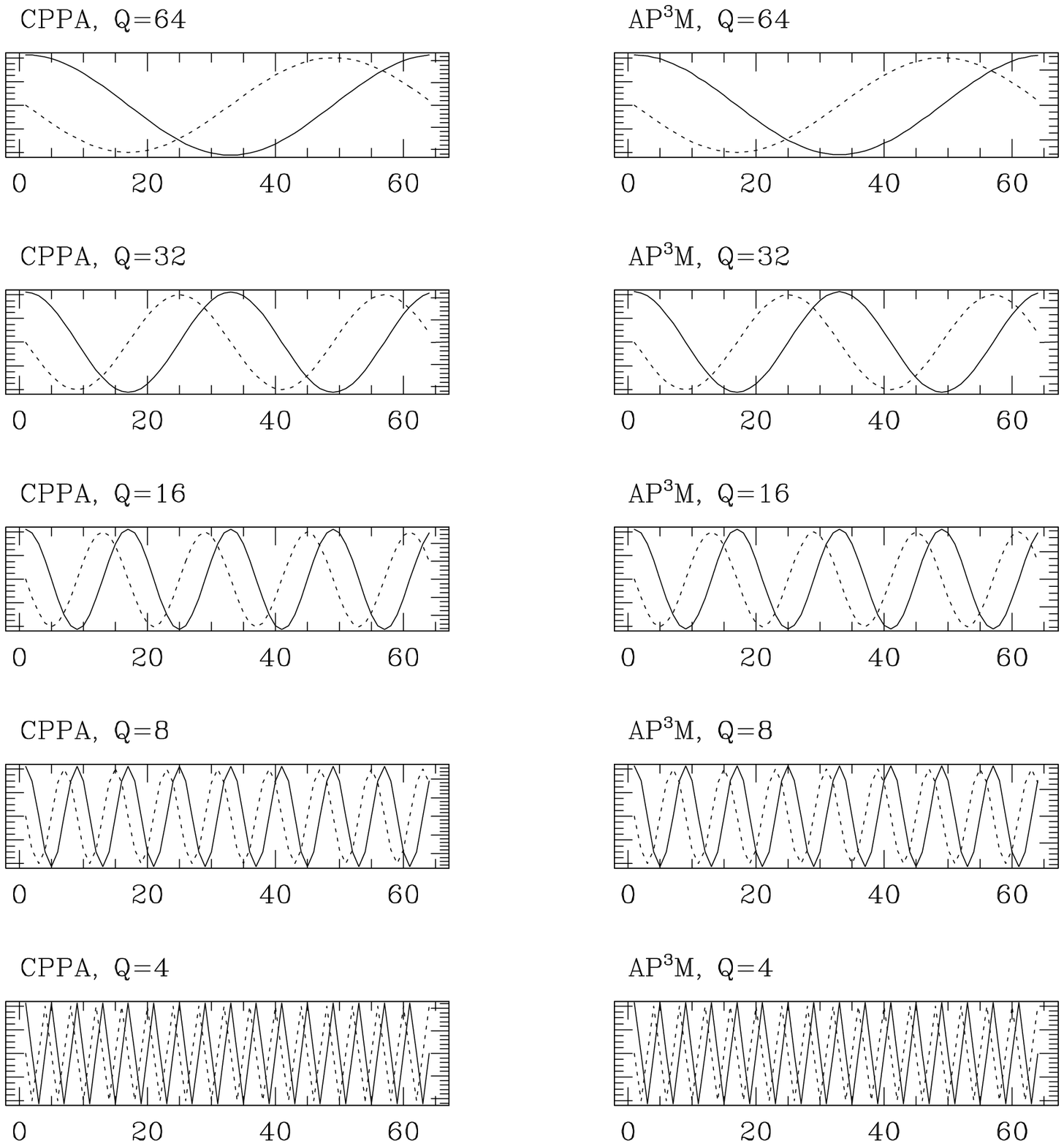}
    \hfill
     }
\FigCap{1-D sections through initial conditions for Zel'dovich pancake test on 
$64^3$ grid.\\ 
Wavelength downwards: 64, 32, 16, 8 and 4 grid cells. {\bf Left:} 
CPPA, {\bf right:} AP$^3$M. {\bf Solid line:} density profile (arbitrary scale) 
{\bf dashed:} velocity profile (arbitrary scale).}
\end{figure} 

\begin{figure}[htb]
    \protect\centerline{
    \hfill
    \epsfysize=14cm\epsffile[5 150 585 700]{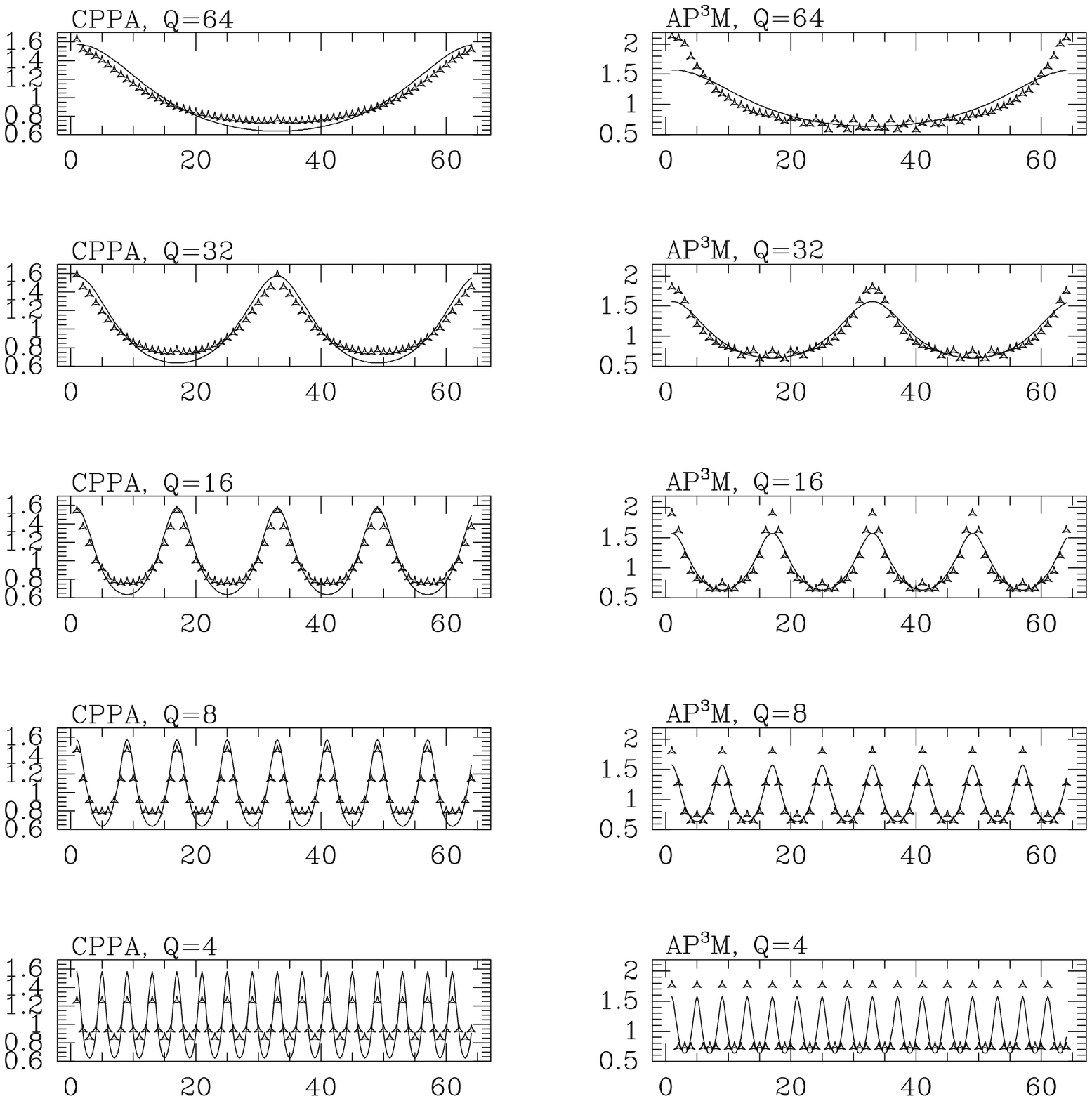}
    \hfill
     }
\FigCap{1-D sections through density profiles for Zel'dovich pancake test on 
$64^3$ grid.\\ 
{\bf Horizontal axis:} position, {\bf vertical axis:} 
$\varrho/\langle\varrho\rangle$. Wavelength downwards: 64, 32, 16, 8 and 4 
grid cells. {\bf Left:} CPPA, {\bf right:} AP$^3$M. {\bf Solid line:} 
analytical results, {\bf points:} simulation.}
\end{figure} 

The density distribution  in the AP$^3$M code is represented by 
the particle coordinates, actually it is a sum of $N$ Dirac delta
functions centered on each of the mass tracers. To obtain
the grid-based density field, we convolved it  with 
a Triangular Shaped Cloud (TSC)  kernel function
$$U(x,y,z) = u(x)u(y)u(z),$$
where 
${u(x)=\frac{3}{4}-x^2}$ for ${|x|\leq\frac{1}{2}}$,
${u(x)=\frac{1}{2}(\frac{3}{2}-|x|)^2}$ for 
${\frac{1}{2}\leq|x|\leq\frac{3}{2}}$,
and $0$ elsewhere (the same kernel is used in the original AP$^3$M).

A good quantitative measure of the departure of the obtained density
field from the predicted theoretical value is the deviation of its
variance.
Integrating $\varrho^2$ over $x$ one can obtain the density dispersion
given by
$$\langle\delta^2\rangle^\frac{1}{2}=
\sqrt{\frac{a_c}{\sqrt{a_c^2-a^2}}-1}.\eqno(19)$$
We obtain our results from a rectangular grid with cell size comparable to the 
regarded wavelengths (formally we are using a cubic top-hat filter). However, 
the predicted analytical value of $\langle\delta^2\rangle^\frac{1}{2}$, 
given by Eq.~(19), ought to be corrected for the effects of finite grid-cell 
size. Suppose our initial perturbation wavelength is $Q$ times larger than a 
grid cell. The value of the predicted density dispersion on the grid is then 
given by 
$$\left.\langle\delta^2\rangle^\frac{1}{2}\right|_{Q}
=\sqrt{\frac{1}{Q}\sum_{p=0}^{Q-1}\left[\frac{Q}{2\pi}\int_\frac{p}{Q}^\frac{p+1}{Q}
\varrho(x,a){\rm d}x\right]^2-1}.\eqno(20)$$
In the weakly nonlinear regime the difference between Eq.~(19) and Eq.~(20) is 
small for long wavelengths, but it becomes important for waves with length of 
several grid cells (see Fig.~3f). 
\begin{figure}[htb]
    \protect\centerline{
    \hfill
    \epsfysize=14cm\epsffile[5 150 585 700]{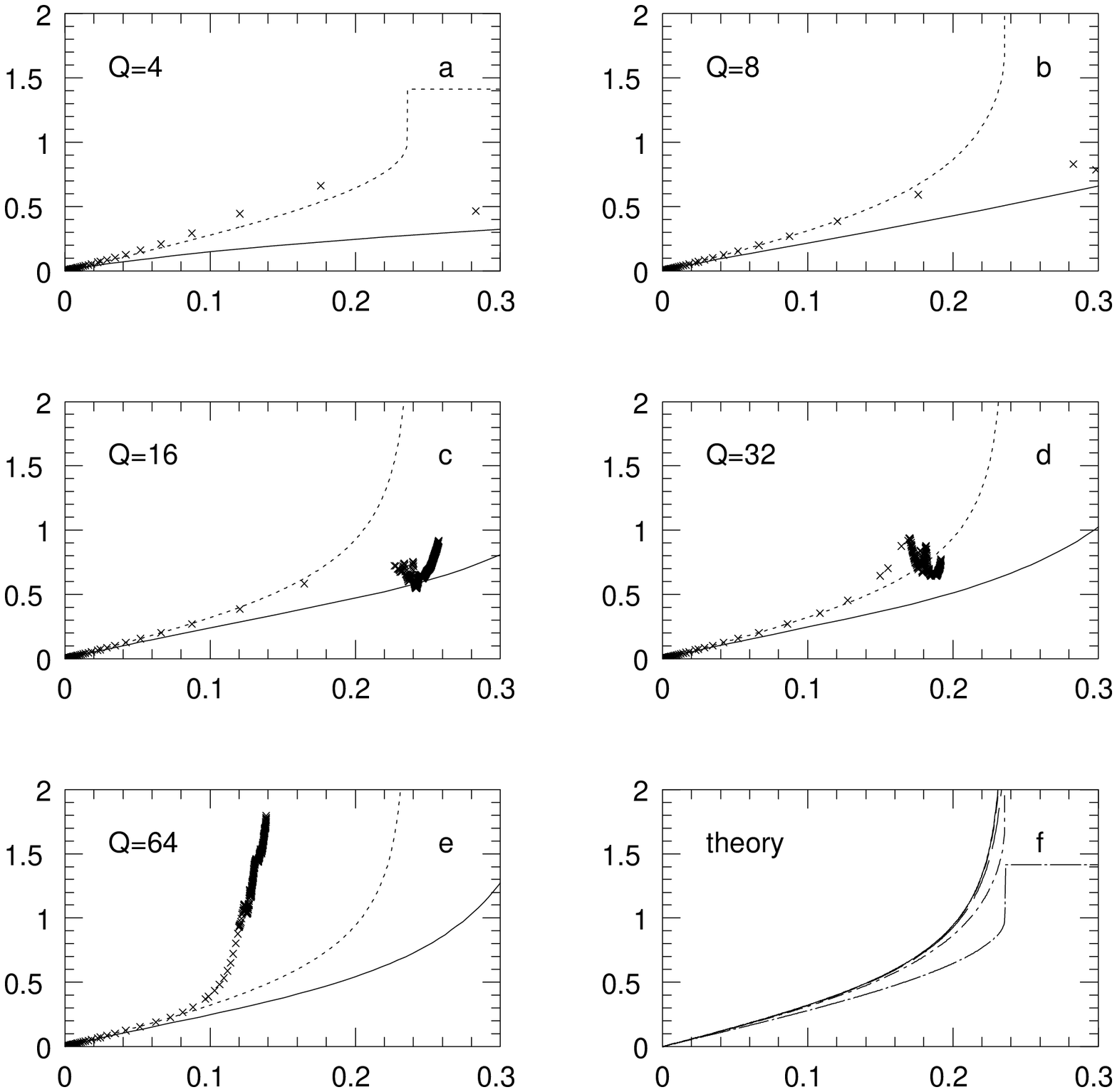}
    \hfill
     }
\FigCap{Evolution of density dispersion in the Zel'dovich pancake test on 
$64^3$ grid.\\ 
{\bf Horizontal axis:} scale factor $a$, {\bf vertical axis:} 
$\langle\delta^2\rangle^{1/2}$ {\bf a}--{\bf e}: simulation results for 
wavelengths of 4, 8, 16, 32 and 64 grid cells. {\bf Solid line:} CPPA, {\bf 
crosses:} AP$^3$M, {\bf dashed line:} theoretical prediction.\\ 
{\bf f}: comparison of theoretical values given by Eq.~(19)
({\bf solid  line}) to Eq.~(20) for ${Q=64}$
${Q=32}$ ({\bf short dash}),
${Q=16}$ ({\bf long dash}), ${Q=8}$ ({\bf dot -- short dash}),
and ${Q=4}$ ({\bf dot -- long dash}).}
\end{figure}

Results of this test for different wavelengths are plotted in Figs~3a--3e. As 
one can see, the evolution of a single wave is qualitatively correct in CPPA, 
but (especially for shorter waves) somewhat slower than predicted. The main 
reason for this is that at the final stage most of the mass is contained 
within walls one cell thick, \ie it has reached the resolution limit. The 
density dispersion curves generated with Couchman's code follow the 
theoretical curve closely until the particles come very close one to another. 
Then the accelerations become very high, the time step gets very short and 
oscillations of the wall due to interpenetration of particle streams may 
start. 

Upon comparing we can state that the CPPA code produces reasonably accurate 
solutions on scales larger than 4 grid cells. 

\Subsection{Scale-free Models}
Three simulations have been performed with each code. In CPPA we used a 
$128^3$ grid, while in AP$^3$M $128^3$ particles on a $128^3$ grid. The models 
were started in an ${\Omega=1}$ universe from Gaussian initial conditions with 
scale-free spectra (power indices +1, ${-1}$ and ${-3}$), cut-off below the 
Nyquist frequency, as explained in Section~2.2. For data reduction the 
density fields were convolved with a Gaussian window of FWHM = 8 cells. 
\begin{figure}[htb]
    \protect\centerline{
    \hfill
    \epsfysize=14cm\epsffile[5 150 585 700]{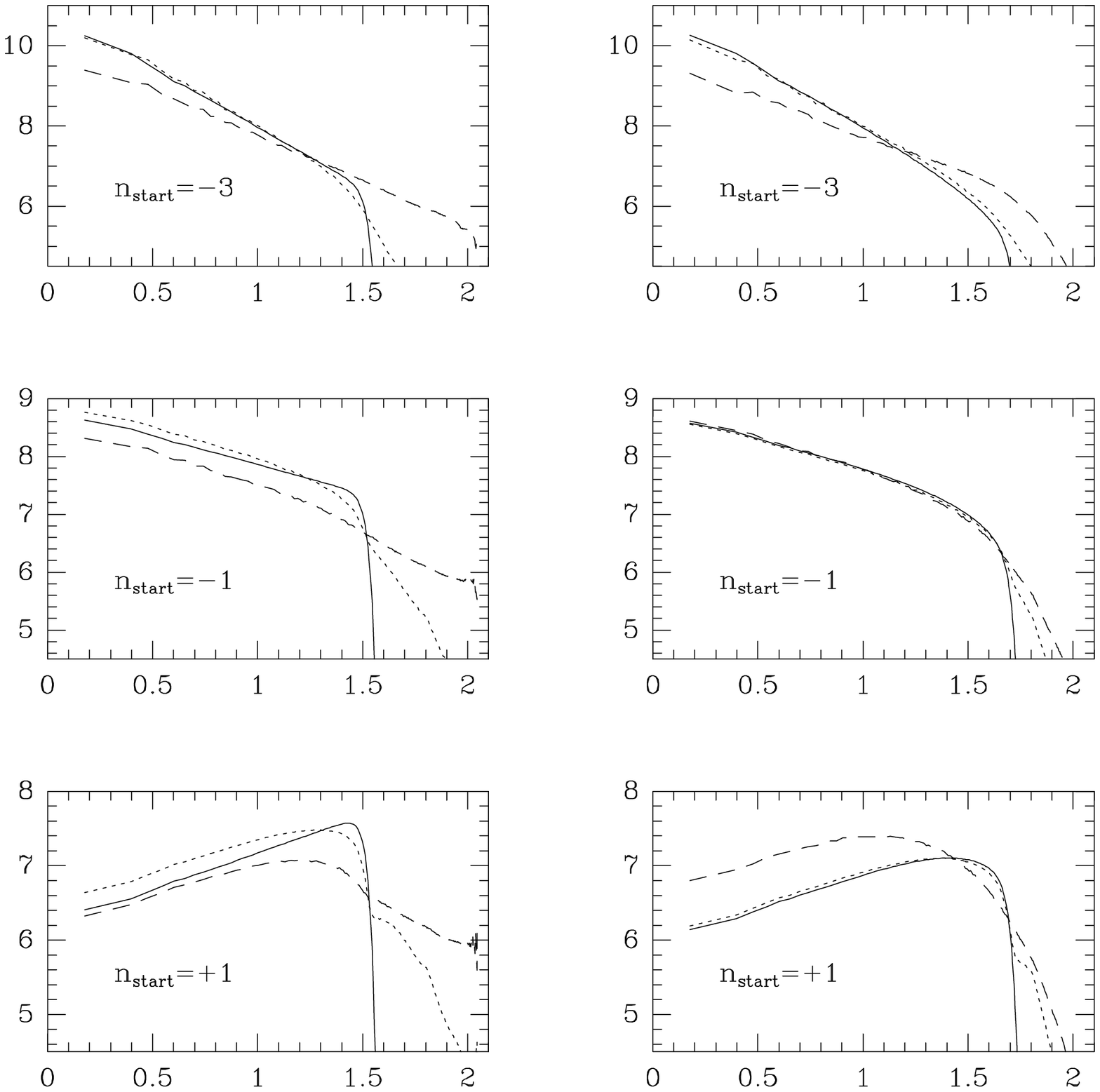}
    \hfill
     }
\FigCap{Evolution of power spectra on $128^3$ grid. {\bf Horizontal axis:} 
$\log k$, {\bf vertical axis:} $\log P(k)$. {\bf Left:} CPPA, {\bf right:} 
AP$^3$M. {\bf Solid line:} initial conditions, {\bf short dash:} weakly 
nonlinear stage (${\langle\delta^2\rangle^{1/2}=0.3}$), {\bf long dash:} 
nonlinear stage (${\langle\delta^2\rangle^{1/2}=2}$).}
\end{figure} 

\begin {figure}[htb]
    \protect\centerline{
    \hfill
    \epsfysize=14cm\epsffile[5 150 585 700]{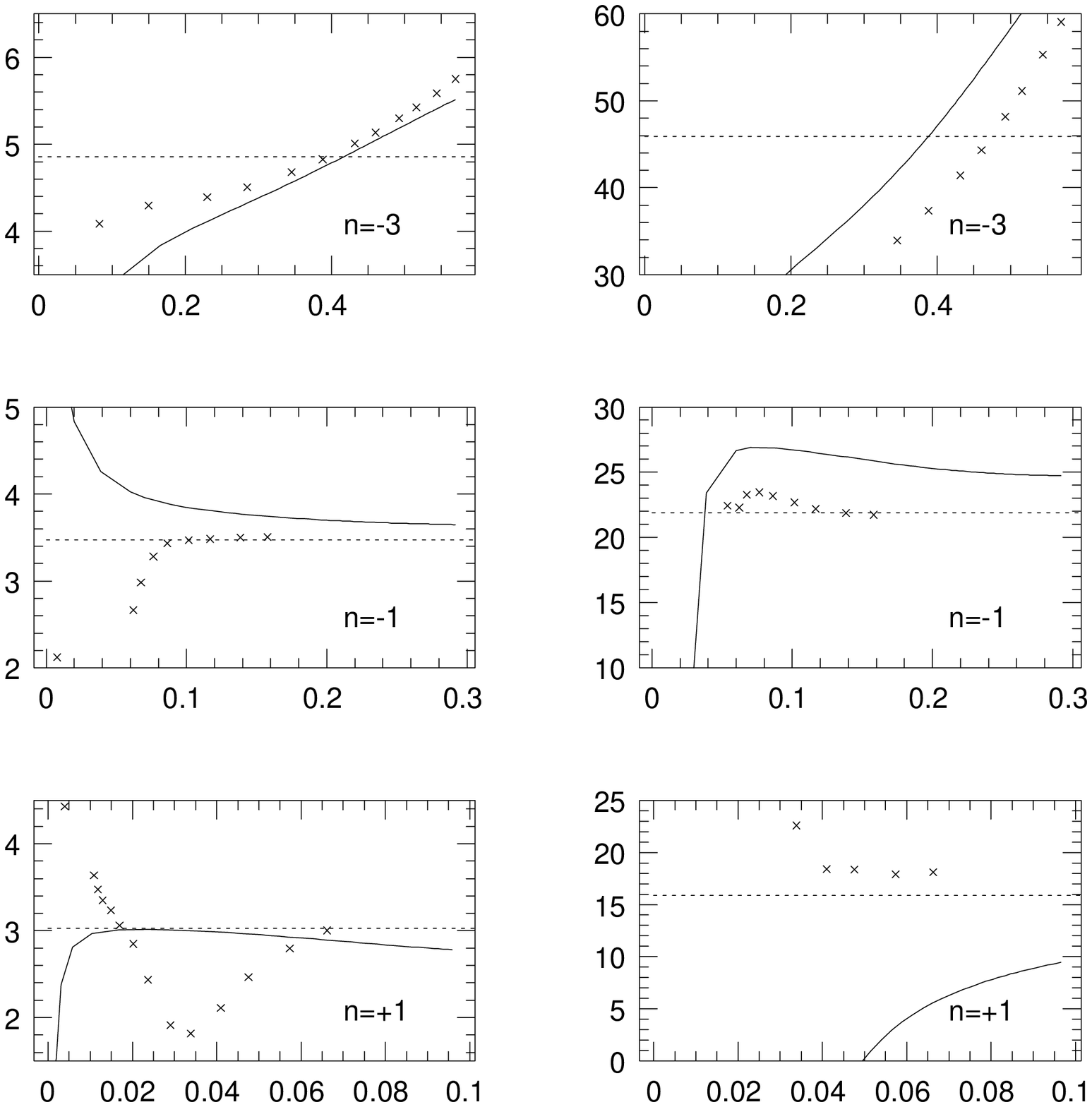}
    \hfill
     }
\FigCap{Evolution of skewness ($S_3$) and kurtosis ($S_4$) of the density 
field versus density dispersion. $128^3$ grid, Gaussian filter. {\bf 
Horizontal axis:} $\langle\delta^2\rangle^{1/2}$ of smoothed density field, 
{\bf vertical axis:} $S_3$ ({\bf left}) and $S_4$ ({\bf right}).\\ 
{\bf Solid line:} PPA, {\bf crosses:} AP$^3$M, {\bf dashed:} perturbation 
theory.}
\end{figure}
The evolution of power spectra is shown in Fig.~4. CPPA as well as AP$^3$M 
conserve the power index on scales larger than 4 grid cells even in the 
nonlinear regime. The evolution of the density field cumulants 
${S_3=\frac{\langle\delta^3\rangle}{\langle\delta^2\rangle^2}}$ and ${S_4= 
\frac{\langle\delta^4\rangle-3\langle\delta^2\rangle^2} 
{\langle\delta^2\rangle^3}}$, compared to exact results obtained with the help 
of perturbation theory (Juszkiewicz, Bouchet, and Colombi 1993, {\L}okas \etal 
1995) is presented in Fig.~5. Results from both codes are in reasonable 
agreement with each other and with theoretical predictions. The largest 
difference is for $S_4$ for ${n=+1}$, however this case often causes 
discrepancies (\eg Catelan and Moscardini (1994) in their Monte-Carlo 
computations obtained a value of ${9\pm1}$, while the perturbation theory 
gives ${S_4=15.9}$). Generally the agreement is better for models with lower 
power index, since in this case there is less power in small scales, on which 
the codes are less accurate. 

\begin{figure}[htb]
%
\FigCap{Cosmic density field for $ \langle\delta^2\rangle^{\frac{1}{2}}=2$. 
Simulation with CPPA code (left) and with AP$^3$M (right). Initial power 
indices: ${n=-3}$ (upper), ${n=-1}$ (middle), ${n=+1}$ (lower).} 
\end{figure} 

Advanced evolutionary stages of the above CPPA and AP$^3$M models are shown in 
Fig.~6 in the form of 2-D slices through the density field. Additionally, we 
performed test runs with identical initial conditions for both CPPA and 
AP$^3$M. The tests were made on a $64^3$ grid with CPPA and with $64^3$ 
particles on a $64^3$ grid in AP$^3$M. The initial spectral indices were ${-
3}$ and ${+1}$. The runs were started at $z=100$, with 
${\langle\delta^2\rangle^{1/2}=0.03}$. The density fields of the evolved 
models are presented in Figs~7 and 8. One can see that all large features 
obtained with both codes are very similar, while small-scale structures 
(important in the ${n=+1}$ runs) are slightly different. Also, the density 
field is smooth in CPPA while in AP$^3$M it is affected by the large scatter 
of the mass tracers in low-density regions. 
\begin{figure}[htb]
%
\FigCap{Cosmic density field for ${n=-3}$ at ${z=2}$. Simulation with AP$^3$M 
(left) and CPPA (right).} 
\end{figure}
\begin{figure}[htb]
%
\FigCap{Cosmic density field for ${n=1}$ at ${z=2}$. Simulation with AP$^3$M 
(left) and CPPA (right).} 
\end{figure}

\Subsection{Performance Comparison}
Timing tests have been performed for both CPPA and AP$^3$M.
The test runs were 60 steps on SUN 10/40 for $32^3$ and $64^3$
particles/cells, with low density contrasts (no mesh refinements in AP$^3$M).
The results are given in Table~1.
\MakeTable{lp{4em}rrrr}{6cm}
{Execution times.}
{
\hline
problem     &     & \multicolumn{4}{c}{execution time [s]} \\
         &     & init  & evol  & total  & per step \\
\hline
 $32^3$ AP$^3$M  &     & 40 & 1129 & 1169 & 18.81  \\
 $32^3$ CPPA  &     & 29  & 371 & 399 & 6.18  \\
         &     &     &     &     &     \\
 $64^3$ AP$^3$M  &     & 261 & 8822 & 9083 & 147.04  \\
 $64^3$ CPPA  &     & 249 & 3987 & 4235 & 66.44  \\
\hline
}
As we can see, CPPA performs one step 2--3 times faster. It has to be mentioned 
that in the non-linear regime in dense regions there are many fast moving 
particles in AP$^3$M, while in CPPA the velocity in a given cell is averaged 
over all mass contained in that cell, thus getting smaller. Hence, due to the 
CFL condition, a step in CPPA is larger than in Couchman's code, making CPPA 
still faster. 

\Section{Summary}
CPPA has proven to be a fast and useful code that well reproduces the 
statistics of cosmic fields in large scales. In the examples presented here it 
lacks resolution on small scales, but presently it is running on large grids 
(up to $192^3$) on SGI and CRAY machines. We are also working on improving its 
accuracy with AMR (Adaptive Mesh Refinement) techniques. 

\Acknow{This work was supported by the KBN grant 2P-304-017-07. The 
simulations are partly performed on Cray computers at the Interdisciplinary 
Center for Mathematical and Computational Modeling in Warsaw.}

\end{document}